\documentclass[prb,twocolumn,floatfix]{revtex4}

\usepackage{color}
\usepackage{ulem}
\usepackage{amsmath}
\usepackage{amssymb}
\usepackage{graphicx}

\begin{document}

\title{Effective theory of quadratic degeneracies}

\author{Y.D.~Chong}
\email{cyd@mit.edu}
\author{Xiao-Gang Wen}
\author{Marin Solja\v{c}i\'{c}}

\affiliation{Department of Physics, Massachusetts Institute of
  Technology, Cambridge, Massachusetts 02139}

\date{\today}

\begin{abstract}
We present an effective theory for the Bloch functions of a
two-dimensional square lattice near a quadratic degeneracy point.  The
degeneracy is protected by the symmetries of the crystal, and breaking
these symmetries can either open a bandgap or split the degeneracy
into a pair of linear degeneracies that are continuable to Dirac
points.  A degeneracy of this type occurs between the second and third
TM bands of a photonic crystal formed by a square lattice of
dielectric rods.  We show that the theory agrees with numerically
computed photonic bandstructures, and yields the correct Chern numbers
induced by parity breaking.
\end{abstract}


\maketitle

\begin{figure}
\centering
\includegraphics[width=0.448\textwidth]{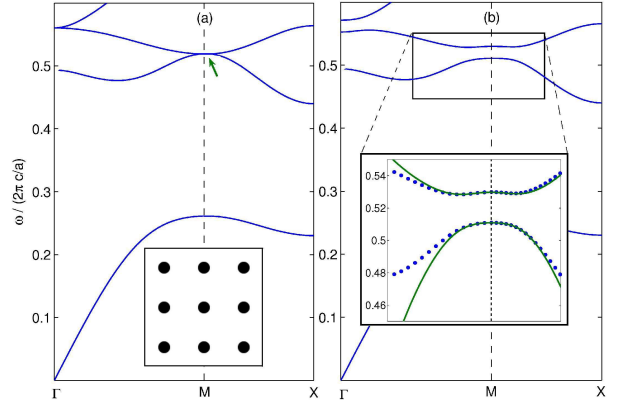}
\caption{(Color online) TM bandstructure of a two-dimensional photonic
  crystal formed by a square lattice of dielectric rods.  (a) Fully
  symmetric crystal with rod radius $0.15a$ (where $a$ is the lattice
  constant) and $\epsilon = 20$. The rods are embedded in air, and
  $\mu = 1$ everywhere.  The quadratic degeneracy between the second
  and third TM bands is indicated with an arrow.  Inset: the crystal
  structure in real space. (b) Parity-broken crystal, with
  off-diagonal permeability component $\mu_{xy} = 0.1i$.  Inset: Band
  structure near the lifted degeneracy, with dots showing numerical
  solutions of the exact Maxwell equations\cite{comsol}.  The solid
  lines show the analytic approximation of Eq.~\ref{symmetric
    eigenvalues}, where $\lambda_0$, $\beta$, and $\gamma$ are
  calculated from the symmetric system and $\alpha_2$ is calculated
  separately based on the proportionality constant with $|\mu_{xy}|$
  (see Fig.~\ref{alphas}.)}
\label{bands}
\end{figure}

In a two-dimensional crystal with a square-lattice ($C_{4v}$)
symmetry, it may happen that a pair of bands are degenerate at a point
of high symmetry, such as the center ($\Gamma$) or corner (M) of the
Brillouin zone.  An example of this occurs in a photonic crystal
formed by a square lattice of dielectric rods.  As shown in
Fig.~\ref{bands}(a), the second and third TM bands exhibit a quadratic
degeneracy at the M-point.  The existence of this degeneracy is
independent of details such as the permittivity and radius of the
rods, as long as the $C_{4v}$ symmetry is preserved.  This degeneracy
is of particular interest since Wang \textit{et.~al.} \cite{WCJS} have
recently shown that it can be lifted by gyromagnetic effects, which
break parity and time-reversal symmetry, and that the bandgap opened
in this way is populated by a one-way edge mode analogous to chiral
electronic edge states in the quantum Hall effect \cite{QHE}.
Electromagnetic one-way edge modes were first predicted by Haldane and
Raghu \cite{Haldane1,Haldane2}, who argued that they typically occur
in systems possessing ``Dirac points'', meaning that the modes near
each degeneracy point can be described by an effective Dirac
Hamiltonian.  Although Dirac points have been extensively analyzed in
the condensed-matter literature \cite{Haldane3}, there has been, to
the best of our knowledge, no analogous study of these quadratic
degeneracies.  Since the system proposed by Wang
\textit{et.~al.}~currently appears to be the most promising for
realizing electromagnetic one-way edge modes, due to its large
relative bandgap \cite{WCJS}, there is a present need for an effective
theory of such degeneracies.

In this paper, we present an effective theory that describes the bands
near a degeneracy point, based on the symmetry properties of $k$-space
around that point.  We show that the quadratic degeneracy in the
$C_{4v}$ crystal can be regarded as a pair of linear degeneracies,
analytically continuable to Dirac points, that are ``pinned'' to the
same $k$-space point by the crystal symmetry.  The quadratic
degeneracy is robust against perturbations that preserve this
symmetry.  It can be lifted by parity and time-reversal symmetry
breaking (which we will henceforth simply refer to as parity
breaking.)  In that case, the two bands acquire Chern numbers
\cite{Haldane1,Haldane2,Simon} of $\pm 1$, in agreement with the
numerical result of Wang \textit{et.~al.}\cite{WCJS} Breaking the
$90^\circ$ rotational symmetry ``unpins'' the quadratic degeneracy
point, which splits apart into two distinct linear degeneracies.  The
theory applies to any two-dimensional Bloch system, whether electronic
or photonic, with $C_{4v}$ symmetry and a quadratic degeneracy point.
In particular, we show that it accurately describes the aforementioned
photonic crystal of dielectric rods for a wide range of dielectric
contrasts and rod radii.

Let us consider a crystal in which two bands are degenerate at a point
$\vec{k} = \vec{k}_M$, with unbroken $C_{4v}$ symmetry.  We choose a
pair of independent Bloch functions at $\vec{k}_M$, denoted by
$u_M^1(\vec{r})$ and $u_M^2(\vec{r})$.  The Bloch functions at
neighboring values of $\vec{k}$ can be written as
\begin{align}
  \begin{aligned}
    u_k^1(\vec{r}) &= c_{11}(\vec{k}) u_M^1(\vec{r})
    + c_{12}(\vec{k}) u_M^2(\vec{r}) \\
    u_k^2(\vec{r}) &= c_{21}(\vec{k}) u_M^1(\vec{r})
    + c_{22}(\vec{k}) u_M^2(\vec{r}).
  \end{aligned}
  \label{bloch decomposition}
\end{align}
The mixing elements $c_{nm}(\vec{k})$ can be related to the mode
frequencies, $\omega_n(\vec{k})$, through an eigenvalue equation
\begin{equation}
  H (\vec{k}) \begin{bmatrix} c_{n1}(\vec{k}) \\ c_{n2}(\vec{k})
  \end{bmatrix} = \lambda_n(\vec{k}) \begin{bmatrix} c_{n1}(\vec{k})
    \\ c_{n2}(\vec{k})
  \end{bmatrix},
  \label{effective theory}
\end{equation}
where the ``effective Hamiltonian'' $H(\vec{k})$ is a $2\times2$
matrix whose eigenvalues are, by definition, $\lambda_n(\vec{k})
\equiv \omega_n(\vec{k}) - \omega_0$.  We will not be concerned with
the value of the ``zero-point'' frequency $\omega_0$.  Now, suppose we
alter the system that we used for defining $u_M^1(\vec{r})$ and
$u_M^2(\vec{r})$, such as by breaking some of its symmetries.  If the
perturbation is sufficiently weak, the Bloch functions of the altered
system can still be described by (\ref{bloch decomposition}) for
$\vec{k} \sim \vec{k}_M$, with some new choice of $c_{nm}(\vec{k})$
and hence of $H(\vec{k})$.

In this system, we will be interested in three different
symmetry-breaking operations.  Firstly, we could ``shear'' the lattice
by rotating the basis vectors as follows:
\begin{align}
  \begin{aligned}
    \vec{a}_1 &= a\left(\cos\theta,\sin\theta\right) \\
    \vec{a}_2 &= a\left(\sin\theta,\cos\theta\right),
  \end{aligned}
  \label{rhombus}
\end{align}
where $a$ denotes the lattice constant.  This breaks the symmetry
under $C_{4}$ rotations and reflections about the $x$ and $y$ axes.
Secondly, we could distort the rods by stretching them along the $x$
or $y$ axes---or, alternatively, stretching the lattice vectors and
rescaling $k_x$ and/or $k_y$; this breaks the symmetry under rotations
and reflections about $y = \pm x$.  Thirdly, we could break parity,
which can be accomplished in a photonic crystal, e.g., using a
magneto-optic effect that adds an imaginary off-diagonal term
$\mu_{xy} = i \eta$ to the permeability tensor of the rods\cite{WCJS}.

The goal is to find an effective Hamiltonian that describes bands such
as those in Fig.~\ref{bands}, including the results of the above
symmetry-breaking operations.  We claim that the desired Hamiltonian
has the following form:
\begin{equation}
  H = \lambda_0 \bigg[ \sum_{i=1}^3 \alpha_i \sigma_i + \beta
    (\kappa_x^2 - \kappa_y^2) \,\sigma_1 + 2 \kappa_x\kappa_y
    \,\sigma_3 + \gamma |\vec{\kappa}|^2 \bigg],
  \label{ansatz}
\end{equation}
where $\vec{\kappa} \equiv \vec{k} - \vec{k}_M$ is the $k$-space
displacement from the degeneracy point and $\sigma_i$ are the usual
Pauli matrices.  The phenomenological parameter $\lambda_0$ determines
the frequency scale, $\beta$ and $\gamma$ control the relative
curvatures of bands along different directions, $\alpha_1$ controls
the relative lengths of the two lattice vectors, $\alpha_2$ is
proportional to the parity-breaking permeability component $\eta$, and
$\alpha_3$ is proportional to the shear angle $\theta$ defined in
(\ref{rhombus}).  This Hamiltonian is valid in the neighborhood of
$\kappa = 0$, and we have omitted $O(\kappa^4)$ terms which have
negligible effects on the band properties in the regime of interest.
Furthermore, we assume that the symmetry-breaking is weak
(e.g.~$\theta \ll 1$), and thus retain only symmetry-breaking terms
that are zeroth-order in $\kappa$.

\begin{figure}
\centering
\includegraphics[width=0.46\textwidth]{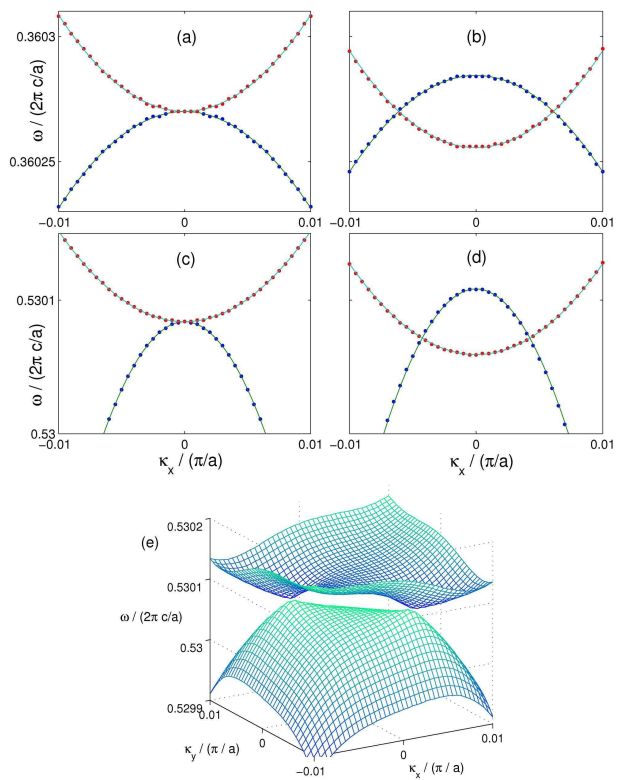}
\caption{(Color online) Photonic bandstructures for time-reversible
  lattices of dielectric rods with radius $r_0$ and permittivity
  $\epsilon$, plotted against the $k$-space displacement from the
  corner of the Brillouin zone, $\vec{\kappa} \equiv \vec{k} -
  (\pi/a,\pi/a)$, along the line $\vec{\kappa} = (\kappa_x, -
  \kappa_x)$.  Dots show numerical data\cite{MPB}; the solid lines
  show the analytic results after fitting the free parameters in
  Eq.~(\ref{ansatz}) to the numerical data.  Note that the
  bandstructure is independent of the parameter $\beta$ along this
  given line.  In (a) and (b), we take $r_0 = 0.25a$ and $\epsilon =
  16.26$, with fitted value $\gamma \simeq 0$.  For (a), $\theta = 0$
  (square lattice) and $\alpha_3 = 0$; for (b), $\theta =
  1.2\times10^{-4}$ radians with fitted value $\alpha_3 =
  3.7\times10^{-5} (\pi/a)^2$.  In (c), (d), and (e), we take $r_0 =
  0.2a$ and $\epsilon = 9.92$, with fitted value $\gamma \simeq -0.5$.
  In (c), $\theta = 0$ and $\alpha_3 = 0$; for (d), $\theta = 10^{-4}$
  radians with fitted value $\alpha_3 = 1.8\times10^{-5} (\pi/a)^2$.
  In (e), a 3\textsc{D} plot of the bandstructure of system (d) is
  shown. }
\label{Dirac points}
\end{figure}

Before discussing the validity of the ansatz (\ref{ansatz}), let us
determine its band structure and then show that it is consistent with
the symmetries of the system.  First, consider $\alpha_1 = \alpha_3 =
0$, for which the eigenvalues are
\begin{equation}
  \lambda_{\pm} (\vec{\kappa})/\lambda_0 = \gamma |\vec{\kappa}|^2 \pm
  \sqrt{|\vec{\kappa}|^4 + (\beta^2 - 1) (\kappa_x^2 - \kappa_y^2)^2 +
    \alpha_2^2}\,.
  \label{symmetric eigenvalues}
\end{equation}
Suppose we assume $\beta = 1$ (setting $\beta \ne 1$ simply distorts
the bandstructure along the $\kappa_x = \pm\kappa_y$ directions).  For
$\alpha_2 = 0$, (\ref{symmetric eigenvalues}) then reduces to a pair
of quadratic bands $\lambda_{\pm}/\lambda_0 = (\gamma \pm 1)
|\vec{\kappa}|^2$, which meet at $\vec{\kappa} = 0$.  The parameter
$\gamma$ controls the relative curvatures of the two bands.  For
instance, when $\gamma = 0$, the bands have equal and opposite
curvatures, as shown in Fig.~\ref{Dirac points}(a); when $-1 < \gamma
< 0$, the two bands curve in opposite directions but the upper band is
flatter, as in Fig.~\ref{Dirac points}(c).  Setting $\alpha_2 \ne 0$
lifts the degeneracy and opens a bandgap $\Delta \lambda =
2\lambda_0\alpha_2$.  The two bands will curve in the \textit{same}
direction at $\vec{\kappa} = 0$, as observed in Fig.~2 of Wang
\textit{et.~al.}  \cite{WCJS} and in Fig.~\ref{bands}(b) of the
current paper.

Next, let us consider $\alpha_3 \ne 0$, keeping $\alpha_1 = 0$.  For $
\alpha_2 = 0$, the quadratic degeneracy splits into two distinct
degeneracy points, at $\vec{\kappa}_\pm = \pm
(\,\sqrt{\alpha_3/2},-\sqrt{\alpha_3/2}\,)$ for $\alpha_3 > 0$ and
$\vec{\kappa}_\pm = \pm (\,\sqrt{-\alpha_3/2},\sqrt{-\alpha_3/2}\,)$
for $\alpha_3 < 0$.  We will henceforth assume that $\alpha_3 > 0$;
the following discussion can be easily adapted to the $\alpha_3 < 0$
case.  Let us expand the Hamiltonian around $\vec{\kappa}_\pm$, using
the variables
\begin{align}
  \begin{aligned}
  q_1 &= \frac{1}{2} \left(\kappa_x + \kappa_y\right) \\ q_2 &=
  \frac{1}{2} \left(-\kappa_x + \kappa_y\right) \pm \sqrt{\alpha_3/2},
  \end{aligned}
  \label{q variables}
\end{align}
which are simply $k$-space displacements from $\vec{\kappa}_\pm$,
rotated by $45^\circ$.  To first order in $q_1$ and $q_2$,
\begin{align}
  \begin{aligned}
  H_{\pm}(\vec{q})/\lambda_0 \; \simeq\; & \gamma (\alpha_3 \mp
  \sqrt{8\alpha_3} \; q_2) \\ & \pm \sqrt{8\alpha_3}\,
    \left(\beta q_1 \sigma_1 + q_2 \sigma_3\right) + \alpha_2 \,\sigma_2.
  \end{aligned}
  \label{Dirac equation}
\end{align}
When $\gamma = 0$ and $\beta = 1$, this reduces to a two-dimensional
Dirac Hamiltonian near each each degeneracy point (or ``Dirac
point'').  Furthermore, $\alpha_2$ plays the role of a mass term,
opening a bandgap $\Delta \lambda = 2\lambda_0\alpha_2$.  Setting
$\gamma \ne 0$ distorts the Dirac Hamiltonian and its eigenvalue
spectrum; for example, the Dirac cones in the $\alpha_2 = 0$ limit
become tilted in $k$-space, as shown in Fig.~\ref{Dirac points}(d).
Along the line $\kappa_y = \pm \kappa_x$, the splitting of the
degeneracy point can be thought of as a vertical relative displacement
of the two parabolic bands (note, however, that the bands meet only at
isolated points in the full $\kappa$-space.)  The splitting is
accompanied by a change in the density of states from a discontinuity
to a linear ``dip'' centered at the frequency of the band degeneracy.
When $\alpha_2 \ne 0$, the density of (TM) states is discontinuous at
the band edges, dropping to zero inside the band gap.

The situation is very similar for $\alpha_1 \ne 0$.  When $\alpha_2 =
\alpha_3 = 0$, the degeneracy splits into two, but along the line
$\kappa_y = 0$ (if $\alpha_1 >0$) or $\kappa_x = 0$ (if $\alpha_1
<0$), instead of $\kappa_x = \pm \kappa_y$.  When both $\alpha_1$ and
$\alpha_3$ are nonzero, the degeneracies are located at an
intermediate location, $\kappa_\pm = \pm (\alpha_1^2 +
\alpha_3^2)^{1/4}(\cos\phi,\sin\phi)$, where $\tan\phi =
\alpha_3/\alpha_1$, and expanding around each point yields a
Dirac-like Hamiltonian analogous to (\ref{Dirac equation}).

When $\alpha_2 \ne 0$, the bands are non-degenerate, and their Chern
numbers \cite{Simon} can be calculated.  The details of this
calculation are given in the Appendix, and the result is that the
upper and lower bands possess Chern numbers $-\mathrm{sgn}(\alpha_2)$
and $\mathrm{sgn}(\alpha_2)$ respectively, regardless of the values of
$\alpha_1$, $\alpha_3$, $\beta$, and $\gamma$.  This implies the
existence of a single family of one-way edge modes \cite{Hatsugai},
and agrees exactly with the numerical results of Wang
\textit{et.~al.}\cite{WCJS}.  Although the effective Hamiltonian
(\ref{ansatz}) is only valid near $\kappa = 0$, it yields the same
Chern number as the actual bandstructure because only the region near
the broken degeneracy point provides a non-vanishing ``Berry flux''
contribution to the Chern number \cite{Simon,Haldane1,Haldane2}.
Furthermore, while our theory only describes weak symmetry-breaking,
the Chern number is a topological quantity and cannot be altered by
non-perturbative distortions (as long as the bands remain
non-degenerate \cite{Simon}), which is why it remains unchanged even
in the strong parity-breaking regime explored by Wang \textit{et.~al.}
When $\alpha_1$ and/or $\alpha_3$ are non-zero, the two linear
degeneracy points each contribute $\pm 1/2$ to the Chern number, in
accordance with previous analyses of the Dirac Hamiltonian
\cite{Haldane3}.  When $\alpha_1 = \alpha_3 = 0$, the Berry connection
winds twice as fast around the point $\kappa = 0$, which provides the
entire contribution of $\pm 1$.  The dependence of the Chern number on
the sign of $\alpha_2$ confirms that $\alpha_2$ controls parity
breaking, since the Chern number can be shown to vanish identically
when parity is unbroken.

The fully symmetric Hamiltonian $H_0 \equiv H|_{\alpha_i = 0}$ must
transform under any operation $g \in C_{4v}$ as
\begin{equation}
  D(g) H_0(\vec{\kappa}) D^{-1}(g) = H_0(g\vec{\kappa}).
  \label{symmetric transform}
\end{equation}
Regardless of the values of $\beta$, $\gamma$, and $\lambda_0$, this
holds if $D(g)$ falls under $E$, the only two-dimensional irreducible
representation of $C_{4v}$.  Thus, $\pm90^\circ$ rotations can be
represented by $\mp i \sigma_2$, reflections about the $\kappa_x$
($\kappa_y$) axes by $\sigma_1$ ($-\sigma_1$), and reflections about
$\kappa_y = \pm \kappa_x$ by $\pm \sigma_3$.

By studying how $H(\vec{\kappa})$ transforms under $E$, we can show
that the quadratic degeneracy is protected by the crystal symmetry.
Any zeroth-order term proportional to the identity matrix, when added
to $H_0$, simply shifts the eigenvalues without opening a gap.  Adding
a zeroth-order term proportional to $\sigma_1$ (i.e. $\alpha_1 \ne 0$)
breaks $C_{4v}$ since, under the representation $E$, $\alpha
\rightarrow - \alpha$ for $90^\circ$ rotations and reflections across
$\kappa_x = \pm \kappa_y$.  Note that $\alpha \rightarrow \alpha$ for
reflections across the $\kappa_x$ and $\kappa_y$ axes, in agreement
with our claim that $\alpha_1 \ne 0$ corresponds to stretching the
lattice.  Similarly, setting $\alpha_2 \ne 0$ preserves the rotational
symmetries but breaks the reflection symmetries (parity).  Finally,
setting $\alpha_3 \ne 0$ preserves the reflection symmetry across
$\kappa_x=\pm \kappa_y$ but breaks the symmetry under $90^\circ$
rotations and reflections across $\kappa_x = 0$ and $\kappa_y = 0$.

Furthermore, the Hamiltonian cannot include terms that are first-order
in $\vec{\kappa}$ if the $C_{4v}$ symmetry is unbroken or only
partially broken.  Such terms have the general form $\Delta H =
\sum_{i=1}^2 \sum_{j=1}^3 \kappa_i c_{ij} \sigma_j$, and we can show
that $c_{ij} = 0$ for all $i,j$ as long as the system is symmetric
under either rotations, reflections about the $\kappa_x$ and
$\kappa_y$ axes, or reflections about $\kappa_x= \pm\kappa_y$.  At
least one of these symmetries is preserved by each of the three
``elementary'' distortions discussed above.  This situation may be
contrasted with a triangular or honeycomb lattice, for which there is
a $C_{3v}$ symmetry around each corner of the hexagonal Brillouin
zone.  There, one can write down an $O(\kappa)$ Hamiltonian which
transforms under a two-dimensional irreducible representation of
$C_{3v}$: this is just the Dirac Hamiltonian
\begin{equation}
  H'(\kappa) = \lambda_0 \left( k_x \sigma_1 + k_y \sigma_3\right).
\end{equation}
In this case, a zeroth-order ``mass'' term proportional to $\sigma_2$
controls parity breaking \cite{Haldane3}.

\begin{figure}
\centering
\includegraphics[width=0.41\textwidth]{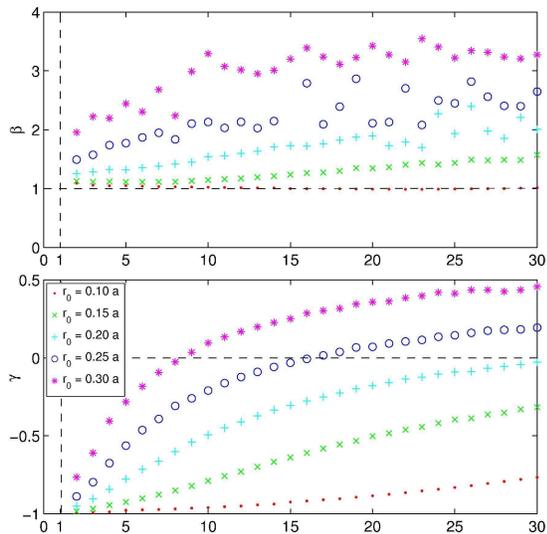}
\caption{(Color online) Values of $\gamma$ and $\beta$ obtained from
  least-squares fits of Eq.~(\ref{symmetric eigenvalues}) to
  bandstructures computed numerically along the lines $k_y = \pi/a$
  and $k_x = -k_y$. The photonic crystal consists of a square lattice
  of rods with radius $r_0$ and permittivity $\epsilon$, embedded in
  air, with $\mu = 1$ everywhere.}
\label{gamma dependence}
\end{figure}

We checked the validity of the ansatz (\ref{ansatz}) and computed the
parameters $\alpha_i$, $\beta$, and $\gamma$ using the MPB \cite{MPB}
and COMSOL \cite{comsol} computer programs, which solve the Maxwell
equations without any approximations apart from the discretization of
the simulation cell.  We first determined $\beta$ and $\gamma$ for
$\alpha_1 = \alpha_2 = \alpha_3 = 0$ by fitting the computed
bandstructures of the fully-symmetric crystal to (\ref{symmetric
  eigenvalues}).  As shown in Fig.~\ref{gamma dependence}, we obtain
values for $\beta$ and $\gamma$ that depend on the details of the
crytal, including the rod radius and permittivity.

\begin{figure}
\centering
\includegraphics[width=0.475\textwidth]{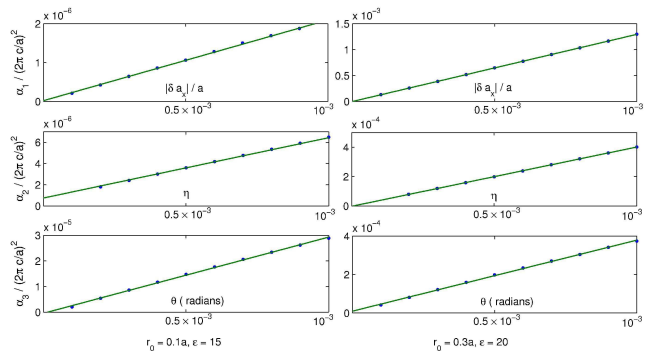}
\caption{(Color online) Plots of $\alpha_1$, $\alpha_2$, and
  $\alpha_3$ for two different photonic crystals.  In the left column,
  the dielectric rods have radius $r_0 = 0.1a$ (where $a$ is the
  lattice constant) and permittivity $\epsilon = 15$.  In the right
  column, $r_0 = 0.3a$ and $\epsilon = 20$.  We obtain $\alpha_1$ from
  the location of the degeneracies $\vec{\kappa} =
  \pm(0,\sqrt{\alpha_1/\beta})$ for lattice stretching parameter
  $\delta a_x / a$.  We obtain $\alpha_2$ by performing a nonlinear
  least-squares fit of Eq.~(\ref{symmetric eigenvalues}) to the
  computed $\mathcal{T}$-broken bandstructures along the line
  $\kappa_x = - \kappa_y$, using the values of $\gamma$ found in
  Fig.~\ref{gamma dependence} (which were obtained using the symmetric
  lattice); it is plotted against the imaginary off-diagonal
  permeability component, $\mu_{xy} = i \eta$.  We obtain $\alpha_3$
  from the location of the degeneracies $\vec{\kappa} =
  \pm(\sqrt{\alpha_3/2},-\sqrt{\alpha_3/2})$, for lattice distortion
  angle $\theta$.}
\label{alphas}
\end{figure}

Upon perturbing the simulated crystals by changing the lattice
constant slightly along (say) the $x$ direction, $\delta a_x < 0$, the
degeneracy splits along the line $\kappa_y = 0$ in the computed
bandstructures.  From the location of the linear degeneracies, we can
obtain $\alpha_1$, which turns out to be proportional to $\delta a_x$.
Similarly, setting $\theta \ne 0$ induces linear degeneracies along
the line $\kappa_x = \kappa_y$, and $\alpha_3$ is found to be
proportional to $\theta$.  By computing the bands for off-diagonal
permeability component $\mu_{xy} = i \eta$ in the rods, and fitting
these to (\ref{symmetric eigenvalues}), we find that $\alpha_2$ is
proportional to $\eta$.  These results are shown in Fig.~\ref{alphas}.
Like $\beta$ and $\gamma$, the proportionality factors
$\alpha_1/\delta a_x$, $\alpha_2/\eta$, and $\alpha_3/\theta$ depend
on the details of the crystal.  The fact that the linear degeneracies
induced by $\alpha_3 \ne 0$ are not determined solely by the lattice
geometry stands in contrast to Dirac points in previously-studied
triangular and hexagonal lattices, which are pinned to the corners of
the Brillouin zone.

In conclusion, we have presented an effective theory for a pair of
bands in the vicinity of a band degeneracy, within a two-dimensional
$C_{4v}$-symmetric crystal.  The high degree of symmetry near the
degeneracy point determines the form of the $2\times2$ effective
Hamiltonian matrix, including the lowest-order symmetry-breaking
terms.  In particular, we showed that quadratic nature of the
degeneracy is protected by the $C_{4v}$ symmetry.  We have also shown
that the theory accurately describes the second and third TM bands of
square-lattice photonic crystals near their M-point degeneracy,
including the correct Chern numbers when parity is broken \cite{WCJS}.
On the other hand, the theory should be applicable to any electronic
or photonic system possessing $C_{4v}$ symmetry and a two-fold band
degeneracy; it is possible, for instance, to construct tight-binding
electronic models that reduce to our effective Hamiltonian
(\ref{ansatz}) near a degeneracy point.  The advantage of the present
method is that it relies only on symmetry principles, and can
therefore be applied to systems, such as photonic crystals, where
other methods such as tight-binding are not applicable.  For lattice
symmetries other than $C_{4v}$, effective Hamiltonians can be
constructed by finding an appropriate representation of the symmetry
group.

\begin{acknowledgments}

We would like to thank Z.~Wang for helpful discussions.  This research
was supported by the Army Research Office through the Institute for
Soldier Nanotechnologies under Contract No.~W911NF-07-D-0004.

\end{acknowledgments}

\appendix*

\section{Calculating the Chern number}

In this Appendix, we describe the calculation of the Berry connection
and Chern number \cite{Simon} for the bands associated with our
effective Hamiltonian.  We will consider the lower band
$|\psi^-(\vec{\kappa})\rangle$; the calculation for the upper band
proceeds analogously.

First, consider $\alpha_1 = \alpha_3 = 0$.  We note that the
eigenvectors of the effective Hamiltonian (\ref{ansatz}) do not depend
on $\gamma$ since that parameter multiplies the identity matrix.  For
simplicity, we set $\beta = 1$.  The eigenvector corresponding to the
lower band is
\begin{multline}
 |\psi^-(\vec{\kappa})\rangle =
 \frac{1}{\sqrt{2[\kappa^4+\alpha_2^2 +
     \kappa^2\sqrt{\kappa^4+\alpha_2^2}\,\sin2\phi]}} \\
 \times \begin{bmatrix}
  -\kappa^2 \cos2\phi + i \beta \\ \kappa^2 \sin2\phi +
  \sqrt{\kappa^4+\alpha_2^2}
 \end{bmatrix},
\end{multline}
regardless of the values of $\gamma$.  Here, $(\kappa,\phi)$ is the
cylindrical coordinate representation of $\vec{\kappa}$.  The Berry
connection is
\begin{align}
\begin{aligned}
  \vec{\mathcal{A}}^{-}(\vec{\kappa}) &=
  \langle\psi^{-}(\vec{\kappa})|\nabla_\kappa|\psi^{-}(\vec{\kappa})\rangle
  \\ &= \frac{i \alpha_2 \kappa \left(\cos2\phi\,\hat{\kappa} -
    \sin2\phi\,\hat{\phi}\right)}{\kappa^4 +
    \beta^2\sqrt{\kappa^4+\alpha_2^2}\,\sin2\phi}.
  \label{berry connection 0}
\end{aligned}
\end{align}
To obtain the Chern number, we integrate the Berry connection around a
loop $\kappa = \kappa_0$:
\begin{align}
  \begin{aligned}
    C^{-} &= \frac{1}{2\pi i} \oint_{\kappa=\kappa_0} d\vec{\kappa}
    \cdot \vec{\mathcal{A}}^{-}(\kappa) \\ &= -
    \frac{2\alpha_2}{\pi}
    \int_{-\frac{\pi}{4}}^{\frac{\pi}{4}} \frac{\kappa_0^2\sin2\phi \;
      d\phi}{\kappa_0^4 + \alpha_2^2 +
      \kappa_0^2\sqrt{\kappa_0^4+\alpha_2^2}\sin2\phi}.
  \end{aligned}
\end{align}
The integral can be performed via the substitution $\sin2\phi = \tanh
u$, and we obtain
\begin{align}
  \begin{aligned}
  C^{-} &= \mathrm{sgn}(\alpha_2) - \frac{\alpha_2}{\sqrt{\kappa_0^4 +
      \alpha_2^2}} \\ &\rightarrow
  \mathrm{sgn}(\alpha_2)\;\mbox{for}\;|\alpha_2| \ll \kappa_0^2.
  \end{aligned}
\end{align}
As discussed in the text, the above result remains unchanged even when
we enter the non-perturbative regime, even though our effective theory
is only valid for small values of $\kappa$ and $\alpha_i$.

When $\alpha_1$ and/or $\alpha_3$ are non-zero, the band maximum at
$\kappa = 0$ splits into two distinct maxima, and expanding around
each maximum yields a Dirac-like Hamiltonian.  For instance, when
$\alpha_1 = 0$ and $\alpha_3 \ne 0$ the maxima occur at
$\vec{\kappa}_\pm = \pm (\,\sqrt{\alpha_3/2},-\sqrt{\alpha_3/2}\,)$,
and the Hamiltonian near each of these points is given by (\ref{Dirac
  equation}).  In terms of the variables $q_1$ and $q_2$ defined in
(\ref{q variables}), the Berry connection for the lower band is
\begin{equation}
  \vec{\mathcal{A}}^{-}_\pm(\vec{q}) = \pm \frac{ib}{2}\cdot
  \frac{\cos\phi \, \hat{q} \,+\, \sin\phi\, \hat{\phi}} {q^2+b^2\pm
    q\sqrt{q^2+b^2}\sin\phi},
  \label{berry connection}
\end{equation}
where $b \equiv \alpha_2 / \sqrt{\alpha_3}$, $\pm$ refers to which
maximum we are expanding around, and $(q,\phi)$ is the cylindrical
coordinate representation of $\vec{q}$.  This Berry connection has the
same form as (\ref{berry connection 0}), but winds half as quickly
around each maximum point as (\ref{berry connection 0}) does around
$\vec{\kappa} = 0$.  Each maximum thus contributes
$\mathrm{sgn}(\alpha_2)/2$ to the Chern number of the lower band.

\end{document}